\documentclass{article}

    \usepackage[final]{neurips_2020}

\usepackage[utf8]{inputenc} % allow utf-8 input
\usepackage[T1]{fontenc}    % use 8-bit T1 fonts
\usepackage{url}            % simple URL typesetting
\usepackage{booktabs}       % professional-quality tables
\usepackage{amsfonts}       % blackboard math symbols
\usepackage{nicefrac}       % compact symbols for 1/2, etc.
\usepackage{microtype}      % microtypography
\usepackage{amssymb}
\usepackage[backend=biber,style=ieee]{biblatex}
\usepackage{graphicx}

\usepackage[breaklinks=true]{hyperref}
\usepackage{mathtools}
\DeclarePairedDelimiter\abs{\lvert}{\rvert}

\title{Tr\"{a}umerAI: Dreaming Music with StyleGAN}

\author{%
  Dasaem Jeong\\
  T-Brain X\\
  SK Telecom\\
  Seoul, South Korea \\
  \texttt{dasaem.jeong@sktbrain.com} \\
   \And
   Seungheon Doh \ \ \  Taegyun Kwon \\
   Graduate School of Culture Technology \\
   KAIST \\
   Dajeon, South Korea \\
   \texttt{\{seungheondoh, ilcobo2\}@kaist.ac.kr} \\
}

\begin{document}

\maketitle

% \begin{abstract}
% Music visualization is a useful tool for enhance music listening experience or delivering musical information in visual way.
% The recent progress in deep learning opened up a new possibility for bridging music and visual art. 

% In this paper, we introduce a neural music visualization based on learned music embedding and StyleGAN. 
% \end{abstract}
\vspace{-3mm}
\section{Introduction}
\vspace{-2mm}
Although music is usually regarded as an audio domain, there are many commonly used visual representations for music, such as music notation, spectrogram, and piano roll. Because the music visualization can provide additional information via visual, there have been various music visualization schemes with different purposes, such as to visualize the emotion of music by selecting photos\cite{chen2008emotion}, to implement an active listening interface by visualizing structure\cite{goto2007active} or progress of music\cite{jeong2016visualizing}, or to create media art performance\cite{taylor2006real}. 
% An experiment also showed that a music visualization could help listeners to memorize the music more precisely \cite{jeong2016visualizing}.

After the recent advances of generative models, there have been several works exploring the cross between music and visual domain using deep neural networks. An audio-reactive StyleGAN\cite{lee2019stylizing} was introduced, which navigate the latent space of StyleGAN with controlled speed based on audio features such as digital filtering outputs and Nsynth\cite{engel2017neural}. 
% The result shows that the generated video reacts to specific sounds such as a bass drum. 
% The limitation is that the selected audio features are focused on the timbral aspects of audio, the video does not match with high-level semantic features of music such as genre and mood. Also, the system controls relative change in latent style space rather than absolute mapping from the audio domain to the visual domain. 
% The limitation is that the images and styles are manually selected to express coherence, and only the movement are controlled by the acoustic features. Also, selected audio features are focused on the timbral aspects of audio, the video does not match with high-level semantic features of music such as genre and mood.
The limitation is that the starting images are manually selected for each generation, rather than automatically generated, and only the movement between images are controlled by the acoustic features. Also, selected audio features are focused on the loudness and timbral aspects of audio, the video does not match with high-level semantic features of music such as genre and mood.

Another recent work proposed a crossing between music and visual style based on artistic periods, such as mapping Debussy's music to French Impressionists' style\cite{lee2020crossing}. The mapping based on the era provides objective shared labels and helps to avoid arbitrariness in pairing music and art. However, as the authors themselves pointed out, the era label is not sufficient enough to bridging between music and paintings. Also, the model generates a visual style rather than an image, so it demands an additional reference image to be style transferred. 

Our goal is to generate a visually appealing video that responds to music with a neural network so that each frame of the video reflects the musical characteristics of the corresponding audio clip. To achieve the goal, we propose a neural music visualizer directly mapping deep music embeddings to style embeddings of StyleGAN, named Tr\"{a}umerAI
% \footnote{Named after Schumann's Tr\"{a}umerei (dreaming), and also dreamer AI in German} 
\footnote{The code is available on 
\href{https://github.com/jdasam/traeumerAI}{https://github.com/jdasam/traeumerAI}. We encourage the readers to watch our demo videos on \href{https://jdasam.github.io/traeumerAI_demo/}{https://jdasam.github.io/traeumerAI\_demo/}}. 

% We manually selected a matching image from the StyleGAN generations for each music clip with 10 seconds length, based on the listener's emotional impression of music and image. Based on 100 pairs of selected music and image, we trained a transfer matrix from the audio embedding domain to the latent style domain. 

% \begin{figure}
%   \centering
%   
%   \label{fig:network}
%   \caption{Main flow}
% \end{figure}

\section{System Implementation}
\vspace{-2mm}
\subsection{Audio Embedding and Image Generator}
\vspace{-1mm}
We utilized a music auto-tagging model as a fixed music encoder, which is a short-chunk CNN with residual connection based on \cite{won2020eval}, and trained the model with MagnaTagATune dataset \cite{law2009evaluation} and its top 50 tags. We used the output of the last CNN layer as an embedding of the audio. 

For image generation, we used StyleGAN \cite{karras2019style} due to its capacity of generating high-resolution images of quality. Our system was made from a public PyTorch implementation\footnote{https://github.com/rosinality/stylegan2-pytorch} of StyleGAN2 \cite{karras2020analyzing} and a pre-trained model that was trained with WikiArt Dataset\footnote{https://github.com/pbaylies/stylegan2}.

\begin{figure}
  \centering
  \includegraphics[width=\textwidth]{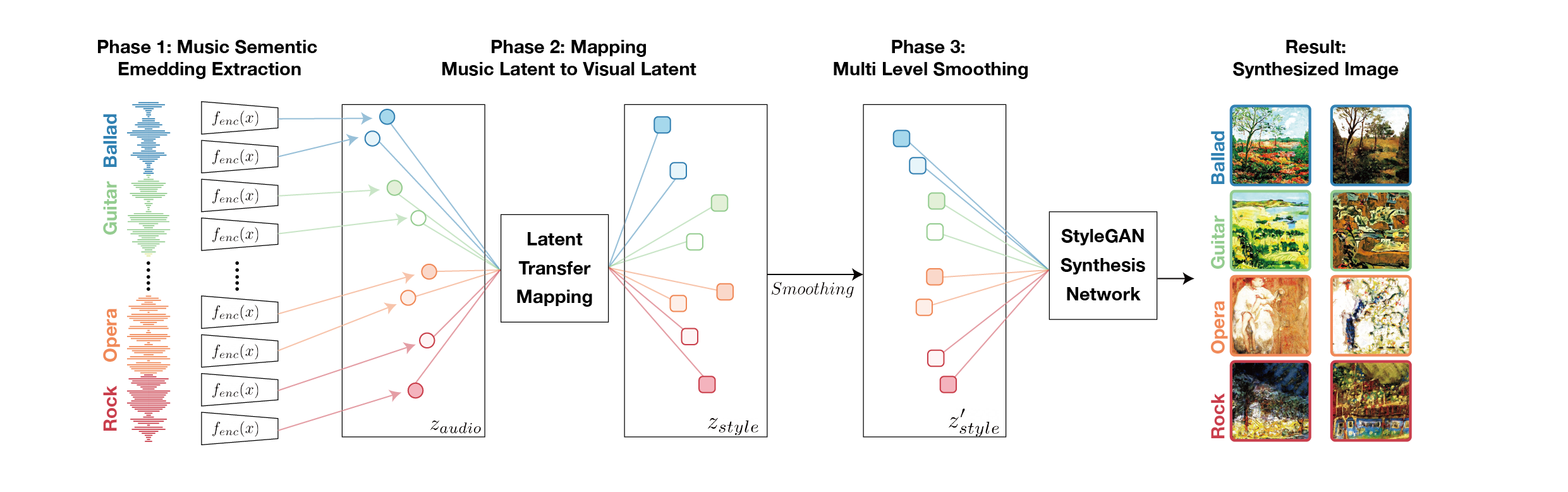}
%   \fbox{\rule[-.5cm]{0cm}{4cm} \rule[-.5cm]{4cm}{0cm}}
  \vspace{-2mm}
  \caption{Structure of the proposed Tr\"{a}umerAI system. Images on the right are generated output of our system from Queen's ``Bohemian Rhapsody''}
  \label{fig:network}

  \vspace{-3mm}
\end{figure}

\subsection{Manual labeling between Music and Image}
\vspace{-1mm}

Rather than establishing an objective metric between musical and visual semantics, we manually labeled the pairs in a subjective manner.
One of the authors listened to 100 music clips of 10 seconds long and selected an image that suits the music among the 200 StyleGAN-generated examples. If the annotator could not find an appropriate image, he generated another 200 images in random or based on a single image from the previous step. The data covers various genres including classical, jazz, pop, ballad, R\&B, new age, K-pop, J-pop, rock, electronic, hip hop, and trot. 

During the process, the annotator considered his emotional impressions such as arousal and valence, genre and era, and timbral characteristic such as instrumentation. We intentionally avoided mapping vocal sounds to a portrait unless the vocal is strongly dominant or any other instruments does not appear, because portraits have relatively low variance on visual style or perceptive emotion.

Based on the collected data, we trained a simple transfer function that converts an audio embedding $z_{mu}$ to a style embedding $w_{st}$. This transfer function is similar to the one used for zero-shot learning between words and images in \cite{socher2013zero}, with mean and deviation $\mu_{st}$, $\sigma_{st}$ of $w_{st}$ from random sampled $z$ of StyleGAN.
\begin{equation}
\mathcal{L}(w_{st}, z_{mu}) = \sum { \abs{w_{st} - (2\sigma_{st} \tanh(W z_{mu} + b) + \mu_{st})}}
\end{equation}
% \mathcal{L}(z_{image}, z_{music}) = \sum{(z_{image} - (W z_{music} + b))^2}

\subsection{Video Generation}
\vspace{-1mm}

The system takes audio input and extracts a sequence of audio embedding. The sequence is converted to a sequence of styles, which is then generated as a sequence of images by StyleGAN. 
During the experiment, linear interpolation between coarsely sampled sequence(~3 sec) results in abrupt acceleration. Therefore, we sampled the 30 audio embeddings per second so that each frame of video is generated from the corresponding audio embedding. Since our mapping conserves the temporal progress of embeddings, the progress of the video followed structural change of music.
% \tg{During the experiment, linear interpolation between coarsely sampled sequence(~3 sec) results in abrupt acceleration. Thus, we sampled the audio embedding every one-third second. Since our mapping conserve the movement of embeddings (with scaled amplitude), the movement of the video followed structural change of music.}
% During the experiment, we found that extracting audio embedding in close interval makes more natural movement in video compared to fill the interval of audio embedding by interpolation. 
Also, smoothing with an averaging window is applied to the style sequence to prevent the generated images from changing too rapidly. The window size differs by style hierarchy, so that coarse, middle, and fine styles are smoothed with a window of 3 sec, 2 sec, and 0.3 sec, respectively.

\section{Discussion}
\vspace{-2mm}
The generated video on Queen's ``Bohemian Rhapsody'', which can be segmented into six different sub-genres, shows that the mapping between audio and video makes a certain level of intra-segment similarity and inter-segment dissimilarity as presented in Figure \ref{fig:network}. 

Although exploring objective mapping between different domains is interesting, subjective mapping can still be a reasonable solution due to the subjective nature of art. Therefore, making a user interface for efficient data labeling between music and painting can be valuable for implementing a personal version of the neural music visualizer. Designing an easily navigable system for the StyleGAN latent space, and applying an active learning method that helps annotators efficiently cover various music or painting styles will significantly reduce the time for the labeling process, which are remained for the future work along with a quantitative evaluation.

% To quantitatively evaluate whether a similar song in the audio domain is paired with a similar image in the style domain, we checked whether ranking in Euclidean distance between audio embeddings has a correlation with a cosine similarity of style embeddings of the paired images. But the average Spearman's rank correlation coefficient was 0.06, while only 11 pairs among 100 pairs showed coefficients with larger than 0.2. This indicates the toughness of finding a bridge between the audio domain to image domain in view of emotional perception. We expect that the mapping would become more consistent if we explore localized and disentangled embedding space.
% This indicates the toughness of finding bridge between audio domain to image domain in view of emotional perception.
% \tg{Did we do this on test set?} \tg{We expect better results if we explore localized and disentangled embedding space, and use meticulous transfer function}

\section*{Ethical Implications}
Since the mapping between music and image is done in subjective pairs, the generated results are heavily biased by the annotator's preference on music. Therefore, the system can generate bizarre or grotesque images from the music, which may deliver a biased impression on the music to viewers, thus may distort the original artist's intention or creativity. 
\printbibliography

\section*{Supplementary}
\subsection*{Example of the Labeling}
\begin{figure}
  \centering
  \includegraphics[width=\textwidth]{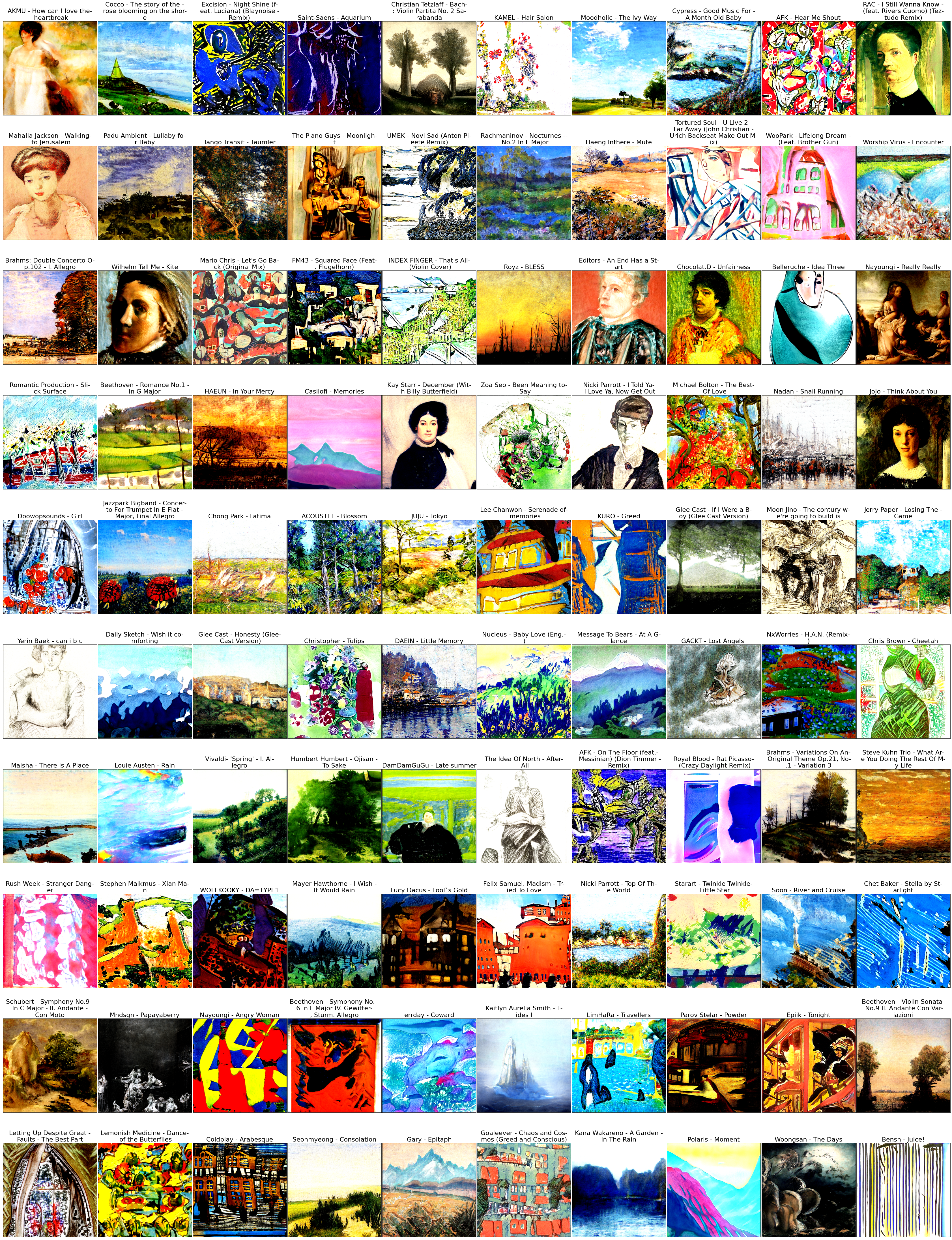}
%   \fbox{\rule[-.5cm]{0cm}{4cm} \rule[-.5cm]{4cm}{0cm}}
  \caption{List of 100 music and corresponding images that were paired during the annotation. Some of the Korean names and titles are translated by the author.}
  \label{fig:images}
\end{figure}

The selected music and corresponding images during the annotation process are presented in Figure \ref{fig:images}. The images were selected from random generation of StyleGAN without truncation to attempt to fully exploit its diversity in expression. Some images like No. 28 and No. 67 are extremely distant from other images in terms of style latent vector, as we did not take into account how extreme the image is in the style latent space. The annotation process took about 10 hours, and we did not modified any annotation after or during the process. To compensate these outliers, we used L1 loss and tanh non-linearity. To be clear, the 100 labeled music clips do not include any music of the artists who are selected for video demonstration.

The annotation process became time consuming when to cover extremely different genre like Korean trot. If an annotator limits the genre of music and style of image in certain level annotation process can become much faster. Therefore, We expect that other users can make their own version of audio-visual mapping by their preference in shorter time.

\subsection*{Example of Generated Images}
\begin{figure}
  \centering
  \includegraphics[width=0.9\textwidth]{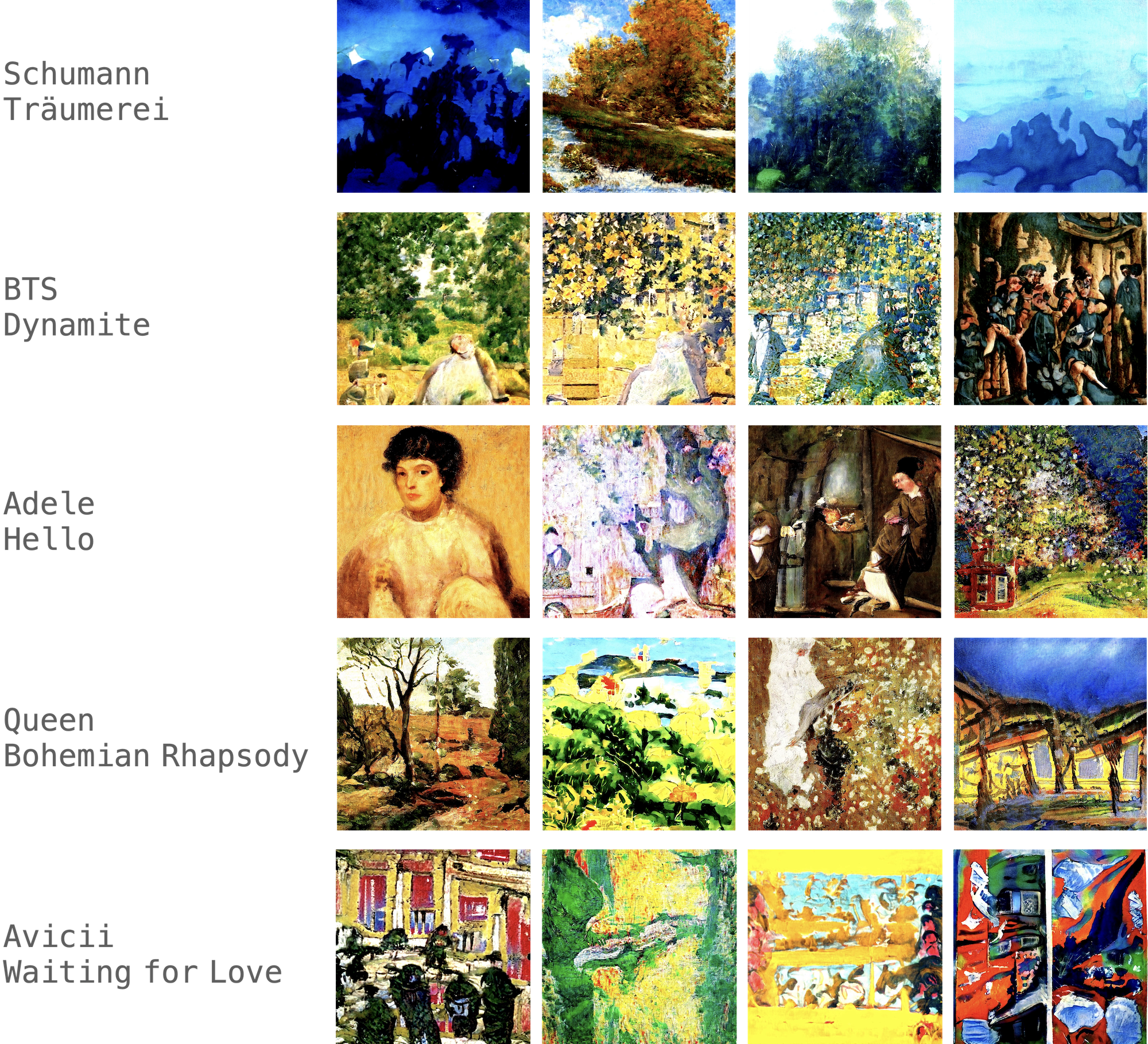}
%   \fbox{\rule[-.5cm]{0cm}{4cm} \rule[-.5cm]{4cm}{0cm}}
  \caption{Snapshots from example videos. Images in the same row are generated from different segments of the same music}
  \label{fig:examples}
\end{figure}

Figure \ref{fig:examples} shows examples of generated images. Again, we encourage the readers to watch videos on \href{https://jdasam.github.io/traeumerAI_demo/}{https://jdasam.github.io/traeumerAI\_demo/}, since how the image is changed throughout the music is the main part of our contribution.

\subsection*{Example of Audio and Style Embedding Trajectory}
\begin{figure}
  \centering
  \includegraphics[width=0.9\textwidth]{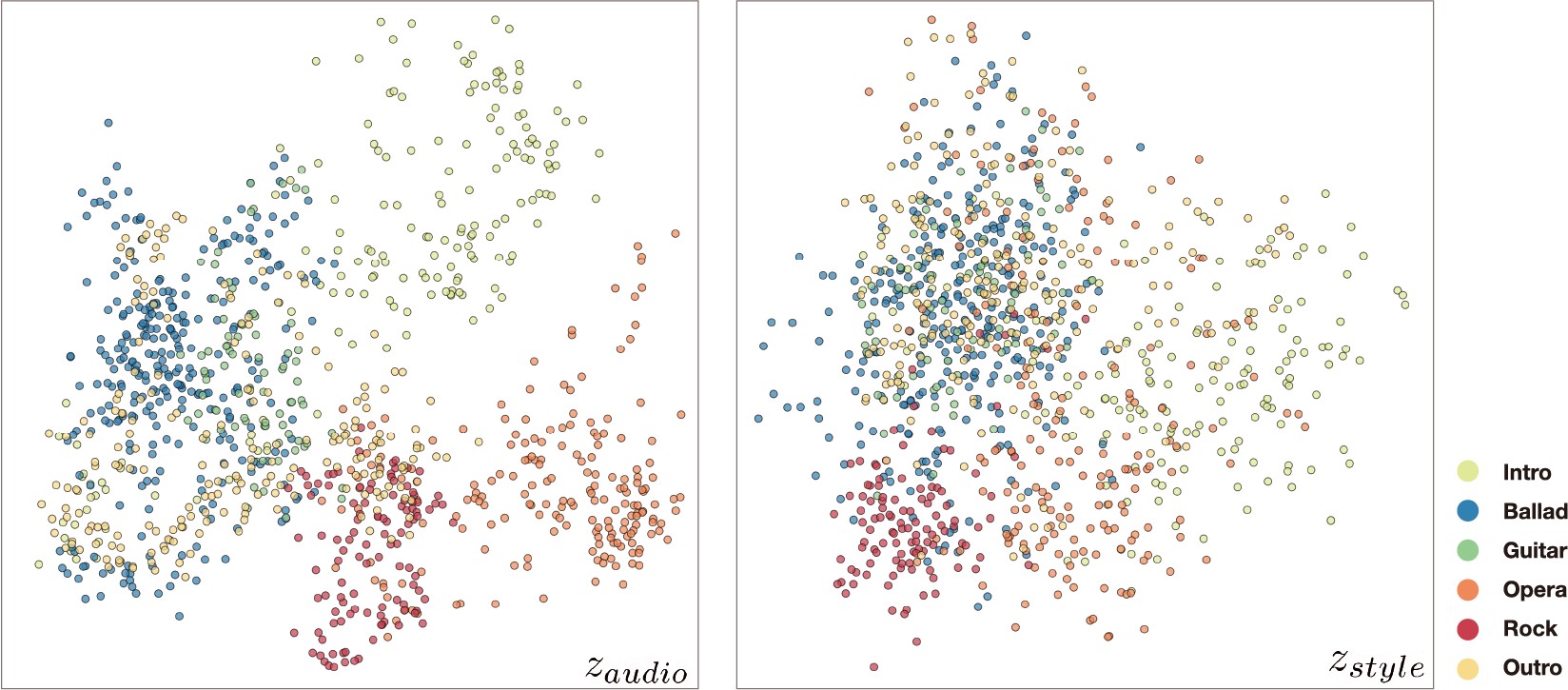}
%   \fbox{\rule[-.5cm]{0cm}{4cm} \rule[-.5cm]{4cm}{0cm}}
  \caption{2D PCA of audio embeddings and mapped style embeddings from Queen's ``Bohemian Rhapsody''. Each point represents 3.7 seconds of audio clip, which is sampled for every one third second}
  \label{fig:trajectory}
\end{figure}
Figure \ref{fig:trajectory} demonstrates how audio embeddings and its corresponding mapped style embeddings changes as music progresses, represented in 2D PCA. The presented result shows that our audio encoder extracts different audio embedding for each sub-genre of the music, and also that the mapped style vectors are conserving the tendency in certain level.

\end{document}